\documentclass[aps,prb,twocolumn,groupedaddress,floatfix]{revtex4}
\usepackage[dvipdf]{graphicx}

\begin{document}
\title{Tight-binding formulation of the dielectric response in semiconductor nanocrystals}
\author{F. \surname{Trani}}
\email{trani@na.infn.it}
\author{D. \surname{Ninno}}
\author{G. \surname{Iadonisi}}
\affiliation{Coherentia CNR-INFM and Dipartimento di Scienze Fisiche, Universit\`a di Napoli Federico II, Complesso Universitario Monte S. Angelo, via Cintia, I-80126 Napoli, Italy}

\begin{abstract}
We report on a theoretical derivation of the electronic dielectric response of semiconductor nanocrystals using a tight-binding framework. Extending to the nanoscale the Hanke and Sham approach [Phys. Rev. B \textbf{12},
4501 (1975)] developed for bulk semiconductors, we show how local field effects can be included in the study of confined systems. A great advantage of this scheme is that of being formulated in terms of localized orbitals and thus it requires very few computational resources and times. Applications to the optical and screening properties of semiconductor nanocrystals are presented here and discussed.
Results concerning the absorption cross section, the static polarizability and the screening function of InAs (direct gap) and Si (indirect gap) nanocrystals compare well to both first principles results and experimental data. We also show that the present scheme allows us to easily go beyond the continuum dielectric model, based on the Clausius-Mossotti equation, which is frequently used to include the nanocrystal surface polarization. Our calculations indicate that the continuum dielectric model, used in conjunction with a size dependent dielectric constant, underestimates the nanocrystal polarizability, leading to exceedingly strong surface polarization fields.
\end{abstract}
\maketitle

\section{Introduction}

The role of the external environment on the electronic properties of semiconductor nanocrystals has been raising a wide interest. This is due to an enhanced control of the synthesis techniques that allows for an optimization of the nanocrystal optical performances. In particular, it has been shown that silicon nanocrystals synthesized in organic solutions have a strong photoluminescence in the blue spectral region, opening the route toward biomedical applications of water soluble silicon nanoparticles.\cite{warner05,jurbergs06,warner05jpc, zhang07}
From the theoretical side, few studies have been done on the influence of the external environment on the semiconductor nanocrystal properties. Indeed, while isolated nanocrystals have been studied in the past using the independent particle approximation,\cite{wang94,allan95,trani05} the role of local field effects (LFEs) has been analyzed within quantum mechanical schemes only recently and for small structures.\cite{bruneval05,gatti05,bruno05} 

A mixture composed of nanocrystals embedded into a dielectric background is usually described using continuum dielectric models (CMs). Such models are based on the assumption that a nanocrystal can be approximated as a continuum medium with a defined position independent static dielectric constant (it can be either the bulk value or a size dependent function), that abruptly goes to the background dielectric constant at the nanocrystal surface. A basic assumption of CMs is that going from the nanocrystal center to the external background, the dielectric constant has a sharp discontinuity across the surface.
In the case of non interacting nanoparticles, these models reduce to the Maxwell-Garnett equations\cite{datta93,ng06,ding05} coupled to the Clausius-Mossotti equation for including the surface polarization fields.\cite{landau} Recent studies have shown that CMs are not always adequate to give a fair description of small nanoparticles. In fact, by looking at the local permittivity profiles, it emerges that close to the surface there is a ``dielectric dead layer,'' and the matching between the dielectric constant inside the structure with that of the background takes place within a finite length\cite{stengel06,delerue03,giustino05,trani06} whose value has a primary importance in the technological applications of semiconductor nanocrystals.
The reduction of the local dielectric constant near the surface is thus the principal cause of the reduction of screening in nanocrystals, whereas the contribution due to the opening of the band gap has a minor role.\cite{delerue03,giustino05,trani07prb}
Moreover, in small and not well shaped nanocrystals, both quantum effects\cite{gatti05} and atomistic features\cite{kim05} become significant, implying that calculations of LFEs within atomistic, quantum mechanical frameworks are definitely required. Unfortunately, the computational cost of this operation rapidly explodes with the nanocrystal size because of the need of computing a large number of excited states. This is the main reason why first principles methods can be used only for very small nanocrystals in a size range that is often significantly below that of the experiments.\cite{ogut03,sottile05} A good compromise between accuracy and the possibility of studying nanocrystals comprising thousands of atoms is obtained by using a real space approach. In the case of dynamic response calculations, time dependent density functional schemes based on a real space, real time approach are the most promising tools for calculating ab initio the optical properties of confined systems with a favorable time scaling.\cite{yabana,castro,andrade,vasiliev}

Our choice in this paper is to use a tight-binding method for determining the dielectric response of confined systems. It is important to stress here that the tight-binding method must be intended in the widest sense and that is the use of a localized basis set. The general framework described in Sec. \ref{sec:framework} and the explicit expressions derived for both the independent particle polarizability and the real space dielectric function can be used with any localized basis set, including the maximally localized Wannier functions.\cite{marzari97} Although the explicit expressions we derived in this work are very general, the applications we have done to illustrate the theory are limited to the simplest version of the tight-binding method where the Hamiltonian matrix elements are fitting parameters determined by the material bulk band structure. We present numerical results for the optical and the screening functions of both silicon and indium arsenide nanocrystals. The good agreement between our results and the experimental data on one side and the results of first principle calculations on the other gives an indication that the chosen route for describing the dielectric response at the nanoscale is very useful. An interesting comparison of the nanocrystal polarizability calculated using the full tight-binding approach and the one obtained using the Clausius-Mossotti equation gives important indications about the range of validity of the dielectric medium theories.

\section{General framework}
\label{sec:framework}
In the following, the linear response theory is reviewed using a localized orbital basis set. Our derivation is based on the framework successfully applied by Hanke and Sham to bulk semiconductors.\cite{hanke75} The main difference consists in the fact that the theory is here developed for confined systems that, as such, do not have periodicity. As a consequence, a different procedure for the calculation of the dielectric constant must be employed. We will show that, neglecting local fields, the standard expression for the nanocrystal dielectric constant within the independent particle random phase approximation (RPA) is retrieved.\cite{wang94,trani05} Instead, an expression is obtained when local fields are taken into account. We refer to the first approximation (neglect of local fields) as RPA, and to the second approximation (inclusion of local fields) as RPA+LF.
Since the main point we wish to address in this paper concerns the role of LFEs in both the optical response and in the screening, we neglect excitonic effects. Besides, while several interesting studies of the excitonic effects in nanocrystals have been done in the past,\cite{leung97,delerue00,lee01} systematic studies on the influence of LFEs are still lacking. The interest in this field is also motivated by studies of semiconductor surfaces done in the past, which showed the important role played by LFEs in that situation.\cite{mochan85,delsole91,wijers91}

\subsection{Tight-binding approach}
As mentioned before, the tight-binding method is a powerful tool for the study of confined systems. Although modern computational facilities and parallel programming lead the calculations to be much more efficient than in the past, first principles studies of nanocrystals containing several thousand atoms are nowadays an almost impossible task. The situation gets even worst if one needs to study the optical response of a nanostructure with the inclusion of local fields, since a calculation of a large number of excited states has to be performed.\cite{sottile05} The tight-binding method has all the advantages of a formulation in terms of localized atomic orbitals whose most important feature is the use of a relatively small basis set, which implies a massive sparsity of the relevant matrices. With appropriate algorithms for the storage, diagonalization, and inversion of a matrix, it is possible to implement an extremely efficient computational tool.

\subsection{Linear response theory}
\label{subsec:linear}

The starting point is, of course, the diagonalization of the single particle Hamiltonian. The nanocrystal wave functions are written as a linear combination of localized atomic orbitals,
\begin{equation}
\label{eq:psi_expansion}
  \psi _n (\mathbf{r}) = \sum _{\sigma \mathbf{R}} B _{\sigma n} (\mathbf{R}) \phi _{\sigma \mathbf{R}} (\mathbf{r}).
\end{equation}
Here, $\sigma$ labels the atomic orbital symmetry and $\mathbf{R}$'s are the atomic coordinates in the nanostructure. The implementation that we use in our calculations is that built on the third-nearest-neighbor $sp^3$ parametrization (with $\sigma = s,p_x,p_y,p_z$) which has been shown to give a good estimation of energy gaps and effective masses of both silicon and indium arsenide, the two materials that we have chosen for illustrating the theory.\cite{niquet00,niquet02,trani04,trani05} 
We neglect the spin-orbit interaction and assume that the basis set is composed by real functions.
In any case, a generalization to include the spin-orbit interaction is straightforward.

The real space independent particle polarizability is defined as \cite{onida02}
\begin{equation}
 \label{eq:polarizability}
  P (\mathbf{r},\mathbf{r'},\omega) = \sum _{\alpha} \left[\frac{f_{\alpha}}{E_{\alpha} - \hbar\omega - \imath \eta}\right]A_{\alpha} (\mathbf{r}) A_{\alpha} (\mathbf{r'}).
\end{equation}
The index $\alpha=(n,n')$ runs over all the possible transitions between the single particle eigenstates. We define $A_{n,n'} (\mathbf{r}) = \psi _n (\mathbf{r}) \psi _{n'} (\mathbf{r})$,  $E_{n,n'} = E_{n'} - E_n$, and $f _{n,n'} = f _{n'} - f _n$. $f _n$ is the $n^{th}$ level occupation number and $\eta$ is a small energy giving rise to a finite broadening of the absorption spectra.
Using Eq. (\ref{eq:psi_expansion}), it is a simple matter to see that
\begin{equation}
 \label{eq:realspacepol} 
 P (\mathbf{r},\mathbf{r'},\omega) = \sum _{\lambda,\mu} P _{\lambda,\mu} \left( \omega\right) A_{\lambda}(\mathbf{r}) A_{\mu}(\mathbf{r'}),
\end{equation}
where $\lambda = (\sigma, \sigma', \mathbf{R})$ labels a pair of atomic orbitals centered on the same site.\cite{hanke74,hanke75} 
In Eq. (\ref{eq:realspacepol}), we have introduced the tight-binding representation of the independent particle polarizability,
\begin{equation}
 \label{eq:tbpolarizability}
 P _{\lambda,\mu} (\omega) = \sum _{\alpha}  \left[\frac{f_{\alpha}}{E_{\alpha} - \hbar\omega - \imath \eta}\right] C _{\lambda \alpha } C _{\mu \alpha}.
\end{equation}
In Eqs. (\ref{eq:realspacepol}) and (\ref{eq:tbpolarizability}), we have defined 
$A _{\lambda} (\mathbf{r}) = \phi _{\sigma \mathbf{R}} (\mathbf{r})  \phi _{\sigma' \mathbf{R}} (\mathbf{r}) $
and $C _{\lambda,\alpha}  = B _{\sigma n} (\mathbf{R}) B _{\sigma' n'} (\mathbf{R})$, respectively.
In order to simplify the calculations, we make a number of
approximations that are strictly related to the parametrization we have used. The first approximation is the neglect of the overlap between atomic orbitals localized on different sites. The very small impact of this assumption has already been checked in the calculation of the optical absorption spectrum of both bulk\cite{tserbak93} and nanocrystalline silicon.\cite{trani05} In both the cases, the agreement between the theory and the experimental data was very good. The second approximation consists in neglecting the off-diagonal intrasite contributions due to atomic orbitals with a different symmetry (terms with $\sigma\neq\sigma'$ in the $\lambda$'s). On this point, there has been a wide discussion in the literature concerning the necessity of introducing additional parameters to take into account these contributions.\cite{graf95,cruz99,foreman02,pedersen01,boykin01,sandu05}
Our experience in this matter is that, apart from the theoretical problem of the breaking of gauge invariance,\cite{foreman02} using a parametrization with many neighbors (i.e., third-nearest-neighbor parametrization) makes the off-diagonal intrasite contributions negligible.\cite{trani05}

The third, most important approximation is in the use of a unique function $A(\mathbf{r})$ for each atomic site, calculated by  averaging all the functions $A_\lambda(\mathbf{r})$  at a given site.
With this approximation, we are neglecting the dependence on the atomic symmetry so that the tight-binding labels reduce to $\lambda=\mathbf{R}$. The averaging is simply given by
\begin{equation}
 \label{eq:Aaverage}
A_{\mathbf{R}}(\mathbf{r}) = \frac{1}{4}\sum _{\sigma=1}^{4} \phi _{\sigma\mathbf{R}}(\mathbf{r}) \phi _{\sigma\mathbf{R}}(\mathbf{r}). 
\end{equation}
Our calculations show that these sets of approximations do not introduce significant errors in the RPA dielectric functions. Within RPA+LF instead, it can induce an error in the calculation of local fields that, according to the quantity one wish to study, can be either neglected or not. For instance, we shall see below that these approximations are responsible of a small difference in the screening when our results are compared with density functional theory results. Nevertheless, it must be stressed that Eq. (\ref{eq:Aaverage}) yields a huge simplification, since it leads to a significant reduction of the size of the matrix to be inverted. In the case of the $sp^3$ parametrization, the tight-binding polarizability matrix $P_{\lambda,\mu}$ is reduced by a factor of $16$.

It is important to underline the role of the Si-H interaction parameters. All the electronic properties can be very sensitive to them, especially in the case of small nanocrystals. For instance, in Si$_{35}$H$_{36}$ the gap energy increases by more than 10\%, with a significant decrease of the static dielectric constant and recombination rates, when the Si-H interactions change from the values of Ref. \onlinecite{niquet00} (obtained by fitting the SiH$_4$ experimental gap) and the parameters of Ref. \onlinecite{delerue06apl} (chosen to give a stronger passivation). Throughout this paper, we use the parameters of Refs. \onlinecite{niquet00} and \onlinecite{niquet02}, which provide a good agreement of the gap energy with other theoretical results and experimental data.

\subsection{Dielectric function}
\label{sec:dielfunc}
The real space dielectric function in the random phase approximation is given by \cite{onida02}
\begin{equation}
\label{eq:epsilon}
  \epsilon (\mathbf{r},\mathbf{r'},\omega) = \delta \left(\mathbf{r}-\mathbf{r'} \right)- \int \mathbf{dr_1} u \left(\mathbf{r},\mathbf{r_1}\right) P \left(\mathbf{r_1},\mathbf{r'},\omega\right),
\end{equation}
where $u$ is the bare Coulomb interaction $u(\mathbf{r},\mathbf{r'}) = e^2/|\mathbf{r'}-\mathbf{r}|$ and the integral is done over the whole space.
Using Eq. (\ref{eq:realspacepol}) for the polarizability, the dielectric function reads
\begin{equation}
  \epsilon (\mathbf{r},\mathbf{r'},\omega) = \delta \left(\mathbf{r}-\mathbf{r'} \right) - \sum _{\lambda, \mu} J _\lambda (\mathbf{r}) P _{\lambda,\mu} \left(\omega\right)A _\mu \left(\mathbf{r'}\right),
\end{equation}
where we have defined the Coulomb integral 
\begin{equation}
\label{eq:coulomb}
  J _\lambda (\mathbf{r}) = \int \mathbf{dr_1} u (\mathbf{r},\mathbf{r_1}) A _\lambda \left(\mathbf{r_1}\right).
\end{equation}

In order to calculate both the optical response and the screening function, the inverse of the real space dielectric function is required. This kind of calculations can be a formidable task if one adopts a first principles point of view.\cite{ogut03} Instead, using the present formulation, the inversion is done
with a modest effort. Using simple matrix properties,\cite{ortuno79,hanke75} the inverse dielectric function $\epsilon ^{-1}$ can be explicitly written as 
\begin{equation}
\label{eq:epsm1}
  \epsilon ^{-1}(\mathbf{r},\mathbf{r'},\omega) = \delta \left(\mathbf{r}-\mathbf{r'} \right) + \sum _{\lambda, \mu} J _\lambda (\mathbf{r}) S _{\lambda,\mu} \left(\omega\right)A _\mu \left(\mathbf{r'}\right).
\end{equation}
In this expression, $S$ is the screened polarizability (also defined as screening matrix\cite{hanke75}) that in the tight-binding representation is
\begin{equation}
\label{eq:Smatrix}
  S _{\lambda,\mu}(\omega) = \sum _{\nu} P_{\lambda,\nu}(\omega) \epsilon _{\nu,\mu}(\omega) ^{-1}.
\end{equation}
where the dielectric function is given by\cite{delerue97}
\begin{equation}
\label{eq:epsilonTB}
   \epsilon _{\lambda,\mu} (\omega) = \delta _{\lambda,\mu} - \sum _{\nu}U_{\lambda,\nu} P_{\nu,\mu} (\omega),
\end{equation}
and the Coulomb interaction matrix is
\begin{equation}
 U _{\lambda,\mu} = \int \mathbf{dr} J _\lambda\left(\mathbf{r}\right) A _\mu\left(\mathbf{r}\right) = e^2\int \mathbf{dr}\mathbf{dr'}  \frac{A _\lambda\left(\mathbf{r}\right) A _\mu\left(\mathbf{r'}\right)}{\left|\mathbf{r}-\mathbf{r'}\right|}.
 \label{eq:coul}
\end{equation}
The main point here is that the inversion of the dielectric function has been
reduced to the inversion of the dielectric matrix defined in Eq. (\ref{eq:epsilonTB}) with enormous advantages with respect to the direct inversion of Eq. (\ref{eq:epsilon}). It is remarkable that the derivation of Eq. (\ref{eq:epsm1}) from Eq. (\ref{eq:epsilon}) is exact, allowing us to reduce the inversion of a large matrix to that of a much smaller one.\cite{ogut03} The physical motivations beyond such a significant reduction of degrees of freedom are rooted on the fact that semiconductor systems can always be studied using few basis functions per atom, as the tight-binding based experience and the Wannier function approaches have widely shown over the years.

Consistent with the approximations outlined in sec. \ref{subsec:linear}, we
parametrize the matrix elements in Eq. (\ref{eq:coul}) as\cite{allan95,delerue97}
\begin{equation}
  U_{\mathbf{R},\mathbf{R'}} = \left\{
\begin{array}{lll}
e^2/\left|\mathbf{R'}-\mathbf{R}\right| &\textrm{ if }& \mathbf{R}\neq\mathbf{R'}\\
U_{at}&\textrm{ if }& \mathbf{R}=\mathbf{R'}.
\end{array}
\right.	
\end{equation}
where $U_{at}$ is an on-site Coulomb interaction term that only depends on the atom located at $\mathbf{R}$. In the applications of the method discussed below, we use the orbital-averaged function $A_{\mathbf{R}}(\mathbf{r})$ defined in Eq. (\ref{eq:Aaverage}) for the calculation of the on-site Coulomb terms $U_{at}$. There is also another procedure for the calculation of the on-site parameters, often used in literature. It consists in the calculation of $U_{at}$ as an average over all the Coulomb interaction energies between pairs of atomic orbitals.\cite{allan95,delerue97,delerue03,niquet02} Below, we shall compare the results obtained with the on-site terms calculated using either the first or the second method.

\section{Dielectric constant}
\label{sec:dielconst}
The dielectric constant is a macroscopic quantity, well defined for extended and periodic systems. However, its meaning in the case of a nanocrystal has been questioned,\cite{delerue03} and new schemes have been proposed for the calculation of a local and position dependent dielectric constant.\cite{delerue03,giustino05} One of the most interesting results in these studies is that well inside a nanocrystal, the local dielectric constant is just the same as that of the corresponding bulk system. However, upon approaching the surface, it rapidly changes until matching the value of the material supporting the nanocrystal. The immediate consequence of this behavior is that the reduction of the dielectric constant in a nanocrystal with respect to the bulk, observed not only theoretically\cite{allan95,wang94,trani05} but even experimentally,\cite{ng06} has to be ascribed to a surface effect\cite{delerue03} which has a dominant weight with respect to the band gap blueshift contribution.\cite{giustino05} It is therefore evident that the surface polarization plays a key role in the dielectric response of confined systems. Within a real space view of the problem, the surface polarization gives a strong discontinuity of the real space polarizability across the nanocrystal surface. 
A key point in this analysis is the validation of the continuum models. These models are based on the idea of studying a nanocrystal as a uniform dielectric sphere. This is a central point when analyzing experimental data simply because it is often not clear what is the real range of validity of these models, particularly when applied to small structures. With our procedure, we are able to calculate local fields within a fully atomistic quantum mechanical scheme, checking, in this way, the CMs upon changing the nanocrystal size.

The macroscopic dielectric constant $\epsilon_M(\omega)$ is defined as the response of a system to a long wavelength macroscopic electric field. In formulas, it is defined as\cite{onida02}
\begin{equation}
 \label{eq:epsilonM}
\epsilon _M (\omega) = \lim _{\mathbf{q}\to\mathbf{0}} \frac{1}{\epsilon^{-1} \left(\mathbf{q},\mathbf{q},\omega\right)},
\end{equation}
where the Fourier transform of the inverse dielectric function is defined as
\begin{equation} 
\label{eq:recepsm1} 
\epsilon ^{-1} \left(\mathbf{q},\mathbf{q'},\omega\right) = \frac{1}{\Omega}\int_{\Omega} \mathbf{dr} \mathbf{dr'} \epsilon^{-1} \left(\mathbf{r},\mathbf{r'},\omega\right) e^{-\imath \mathbf{q'}\mathbf{r'}+\imath\mathbf{q}\mathbf{r}}.
\end{equation}
We calculate the Fourier transform of real space quantities, localized on a finite structure, approaching a constant value outside the structure. In particular, the dielectric function defined in Eq. (\ref{eq:epsilonM}) is equal to the dielectric constant of the background embedding medium in the region of   space outside the nanocrystal. Here, we consider spherical nanocrystals in vacuum, so $\epsilon (\mathbf{r},\mathbf{r'},\omega) = 1$ when $|\mathbf{r}|,|\mathbf{r'}|>R$ ($R$ is the radius, and the coordinate system is placed at the center of the nanocrystal).

In order to make spatial integrations, we introduce an integration volume $\Omega$. By using values of $\Omega$ much greater than the nanocrystal volume (taking the limit $\Omega\to \infty$), we retrieve $\Omega$-independent results. A main point is that the dielectric constant calculated by Eq. (\ref{eq:epsilonM}) is not just the nanocrystal dielectric constant, since it includes the vacuum space encompassing the nanocrystal. Defining the filling factor f=$\Omega_S/\Omega$ as the ratio between the nanocrystal volume $\Omega_S$ and the integration volume $\Omega$, we can write the dielectric constant defined in Eq. (\ref{eq:epsilonM}) as the average between the nanocrystal dielectric constant $\epsilon _{S}$ and that of the vacuum space ($\epsilon _{out} = 1$),
\begin{equation}
\label{eq:meanfield}
\epsilon _{M} = (1-f) + f\epsilon _{S}.
\end{equation}
From Eq. (\ref{eq:meanfield}), the nanocrystal dielectric constant $\epsilon_S$ can be calculated. 

By substitution of Eq. (\ref{eq:epsilon}) into Eq. (\ref{eq:recepsm1}), the following expression is obtained:
\begin{equation}
 \epsilon ^{-1} \left(\mathbf{q},\mathbf{q},\omega\right) = 1 + \frac{u (\mathbf{q})}{\Omega}  \sum _{\lambda, \mu} A _\lambda (\mathbf{q}) S _{\lambda,\mu} \left(\omega\right)A _\mu \left(\mathbf{q}\right),
\end{equation}
where the Coulomb interaction takes the usual form $u (\mathbf{q}) = 4\pi e^2/q^2$ and $A(\mathbf{q})$ is the Fourier transform of $A (\mathbf{r})$. In the $\mathbf{q}\to 0$ limit, it can be expanded using the dipole approximation  
\begin{equation}
   A _{\lambda} (\mathbf{q}) \simeq 1 - \imath  \mathbf{q} \cdot \mathbf{D}_\lambda,
\end{equation}
where the dipole matrix elements in the tight-binding basis set are given by $D^{\alpha}_{\lambda} = \int \mathbf{dr} \left[x_{\alpha} A_{\lambda} (\mathbf{r})\right]$, $\alpha$ being the Cartesian component $x$, $y$ or $z$.
The dielectric constant defined by Eq. (\ref{eq:epsilonM}) is thus
\begin{equation}
  \epsilon ^{\alpha}_{M}  \left(\omega\right) = \left[1 + \frac{4\pi e^2}{\Omega} \sum _{\lambda,\mu} D^{\alpha}_{\lambda}S _{\lambda,\mu} \left(\omega\right) D^{\alpha}_{\mu} \right]^{-1}.
\end{equation}
In order to simplify the notation, in the following we do not write the explicit dependence on the Cartesian component $\alpha$. Since in the examples given below we only consider spherical nanocrystal of cubic semiconductors, the dielectric tensor reduces to a constant, and the dependence on $\alpha$ disappears. However, the result can be easily generalized to anisotropic systems.
In the limit $\Omega \to\infty$, the previous expression becomes 
$\epsilon _{M}  \left(\omega\right) = 1 - 4\pi e^2 \sum _{\lambda,\mu} D_{\lambda}S _{\lambda,\mu} \left(\omega\right) D_{\mu} /\Omega $, and according to Eq. (\ref{eq:meanfield}), the nanocrystal dielectric constant can be finally derived as
\begin{equation}
\label{eq:epsnclf}
\epsilon _{S} \left(\omega\right) = 1 - \frac{4\pi e^2}{\Omega_S} \sum _{\lambda,\mu} D_{\lambda} S_{\lambda,\mu} \left(\omega\right) D_{\mu}.
\end{equation}
As expected, the integration volume simplifies in all expressions and only the nanocrystal volume appears in the final expression.
We point out that this is the dielectric constant of a confined structure, defined in linear response theory within RPA with LFEs included (RPA+LF scheme).  If we neglect local field effects (RPA), we simply take the $\mathbf{q}\to0$ limit of the reciprocal space dielectric function,
\begin{equation}
\epsilon _M ^{RPA}(\omega) = \lim _{\mathbf{q}\to\mathbf{0}} \epsilon \left(\mathbf{q},\mathbf{q},\omega\right),
\end{equation}
retrieving a standard expression for the RPA dielectric constant,
\begin{equation}
\label{eq:epsncrpa}  
  \epsilon _{S}  \left(\omega\right) = 1 - \frac{4\pi e^2}{\Omega_S} \sum _{\lambda,\mu} D_{\lambda}P_{\lambda,\mu}\left(\omega\right) D_{\mu}.
\end{equation}

The difference between the dielectric function with and without LFEs is in the polarizability term that is either screened as in Eq. (\ref{eq:epsnclf}) or unscreened as in Eq. (\ref{eq:epsncrpa}). We refer the reader to Ref. \onlinecite{trani07prb} for a detailed discussion on the physical interpretation of the difference between the dielectric constant calculated within both the schemes and on the role played by local fields.

\subsection*{Results}
\begin{figure}
 \includegraphics[height=0.35\textheight]{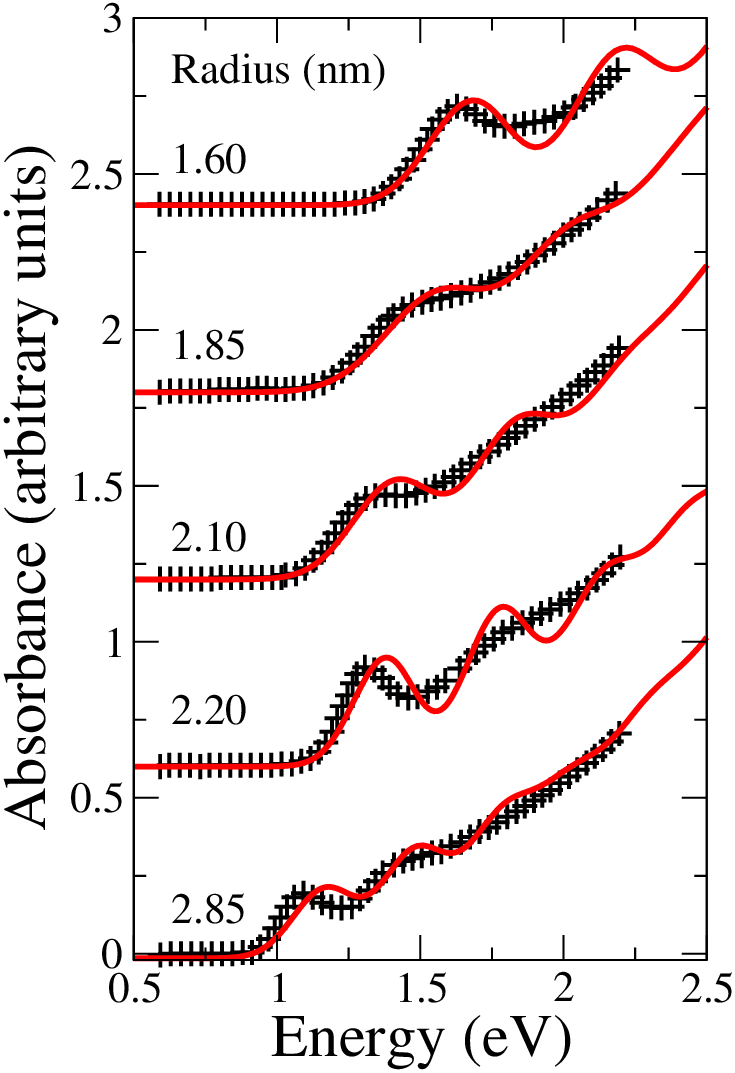}
 \caption{ \label{fig:fig1} (Color online)
Absorption spectra of colloidal InAs nanocrystals. Present tight-binding calculations (red lines) are compared to experimental data of Ref. \onlinecite{yu05} (black symbols). A Gaussian broadening has been used.}
\end{figure}

In this section, the optical properties of Si and InAs nanocrystals are illustrated. We recently applied the method to the calculation of the Si nanocrystal absorption spectra, and good agreement with experiments was recovered.\cite{trani07prb} We report in Fig. \ref{fig:fig1} the absorption spectra of InAs nanocrystals calculated with the present RPA+LF tight-binding scheme, and compare them with experimental data taken from recent measurements on colloidal nanocrystals.\cite{yu05} We use a second-nearest-neighbor tight-binding parametrization,\cite{niquet02} calculated by fitting the band structure and the effective masses of bulk InAs. As Fig. \ref{fig:fig1} shows, the agreement between theory and experiments is good.

\begin{figure}
\includegraphics[width=0.45\textwidth]{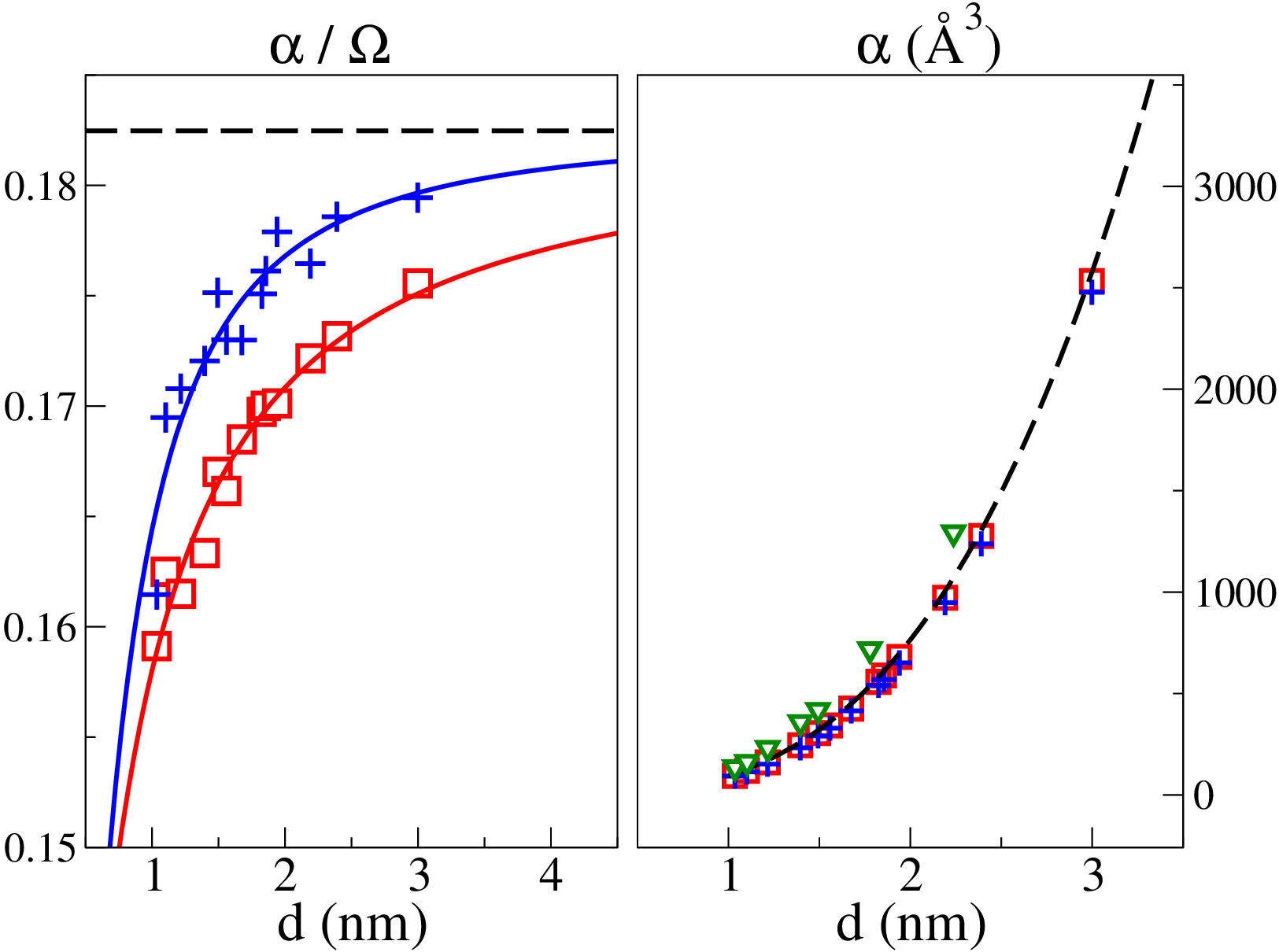}
 \caption{\label{fig:fig2}
(Color online) Spherical nanocrystalline Si polarizability per volume (left panel) and polarizability (right panel) as a function of the nanocrystal size. Red squares symbols, RPA with the continuum dielectric model; blue cross symbols, RPA+LF; dashed line, bulk limit. The solid lines are curves obtained by interpolation (see text), with the following fitting parameters: RPA, $\alpha=3.5$ \AA{} and $l= 1.18$, RPA+LF, $\alpha=3.25$ \AA{} and $l=1.77$. The green triangle symbols are the TDLDA results obtained in Ref. \onlinecite{tsolakidis05}. }
\end{figure}

In Fig. \ref{fig:fig2}, the static polarizability per volume (left panel) and the static polarizability (right panel) of spherical Si nanocrystals are shown. Classically, the polarizability of an atomic cluster can be calculated starting from the bond polarizabilities and including the dipole-dipole interactions in order to take into account local field effects.\cite{kim05}
Within a RPA+LF scheme, however, local fields are already included in the calculation of the dielectric constant, and thus the nanocrystal polarizability is directly related to it through the expression
\begin{equation}
 \alpha=\frac{\Omega_S}{4\pi}\left(\epsilon_S - 1\right)  \qquad \mathrm{(RPA+LF)}
\end{equation}
The RPA+LF results are indicated in Fig. \ref{fig:fig2} with blue cross symbols.
An alternative way of calculating the polarizability is by modeling the nanocrystal as a continuum dielectric system. Within this model, the calculation of the surface polarization effects leads to the Clausius-Mossotti expression.\cite{kim05} This approach is often used by modeling the nanocrystal either with the bulk Si or a size dependent dielectric constant. In order to check whether the continuum model works well for nanocrystals, we use the RPA static dielectric constant, and calculate the polarizability through the Clausius-Mossotti expression\cite{tsolakidis05}
\begin{equation}
\label{eq:clausius}
 \alpha=\frac{3\Omega_S}{4\pi} \left[\frac{\epsilon_S - 1}{\epsilon_S +2}\right] \qquad \mathrm{(RPA)}.
\end{equation}
The results obtained with this approximation are shown by red square symbols. In the limit of large nanocrystals, the RPA dielectric constant converges to the bulk value,\cite{trani05} and we retrieve a large size limit for the polarizability per volume, indicated with a black dashed line in the left panel of Fig. \ref{fig:fig2}. This value has been obtained using the experimental bulk Si dielectric constant in Eq. (\ref{eq:clausius}).\cite{note} 
With the use of an interpolation formula for the size dependent dielectric constant\cite{wang94},
\begin{equation}
 \epsilon(R) = 1 + \frac{\epsilon_b - 1 }{ 1 + (\alpha / R)^l}
\end{equation}
a fit of the present tight-binding results is obtained. The results are shown as full lines in Fig. \ref{fig:fig2}. It can be noted that the RPA results combined with Eq. (\ref{eq:clausius}) underestimate the polarizability per volume, in agreement with our previous dielectric constant calculations.\cite{trani07prb} On the right panel of Fig. \ref{fig:fig2}, the tight-binding results for the polarizability are compared to a first principles time dependent local density approximation (TDLDA) calculation.\cite{tsolakidis05} The agreement for small nanocrystals is very good. However, upon increasing the nanocrystal size, there is a discrepancy between TDLDA and tight-binding results. This is probably due to the overestimation of the bulk Si dielectric constant within LDA which leads to an overestimation of the static polarizability, with an error that increases with the nanocrystal size.\cite{tsolakidis05}

It is worth trying to better understand the left panel of Fig. \ref{fig:fig2}. From the figure, it emerges that the nanocrystal polarizability per volume is always smaller with respect to the large size limit. As the recent literature has shown, this decrease of the polarizability for nanosized objects is mostly due to an overall decrease of the local polarizability across the surface, the quantum confinement effect giving a relatively small contribution.\cite{delerue03,giustino05,shi06}
It is extremely interesting even from an experimental point of view trying to understand the origin of the differences shown in Fig. \ref{fig:fig2} between the full calculation and the continuum dielectric model.
There are at least three sources of errors. First, the Clausius-Mossotti equation assumes that the local field inside the structure is position independent, whereas even classical atomistic calculations have shown that the surface bonds see a local field very different with respect to the local field seen by the bonds placed inside the nanocrystal.\cite{kim05} A second difference is that the surface bonds have a different polarizability with respect to the bulk Si.\cite{delerue03,giustino05,shi06} Third, the full calculation takes into account the quantum confinement effect.

All of these effects can give opposite contributions to the polarizability, and it is not obvious whether the CMs either overestimate or underestimate the correct results. In light of these considerations, our results appear significant since they state that the full results lie somehow in the middle between the CMs used with either a size dependent or the bulk limit static dielectric constant. The relative error made in using the CMs is of the order of 2\%-5\% in the size range considered here.

A last significant finding is that the polarizability per volume calculated within RPA+LF is almost independent of the choice of the on-site parameter U$_{at}$, showing a robustness of the method versus U$_{at}$ in the polarizability calculations.

\section{Point charge screening}
\label{sec:screening}
\begin{figure}
\includegraphics[width=0.45\textwidth]{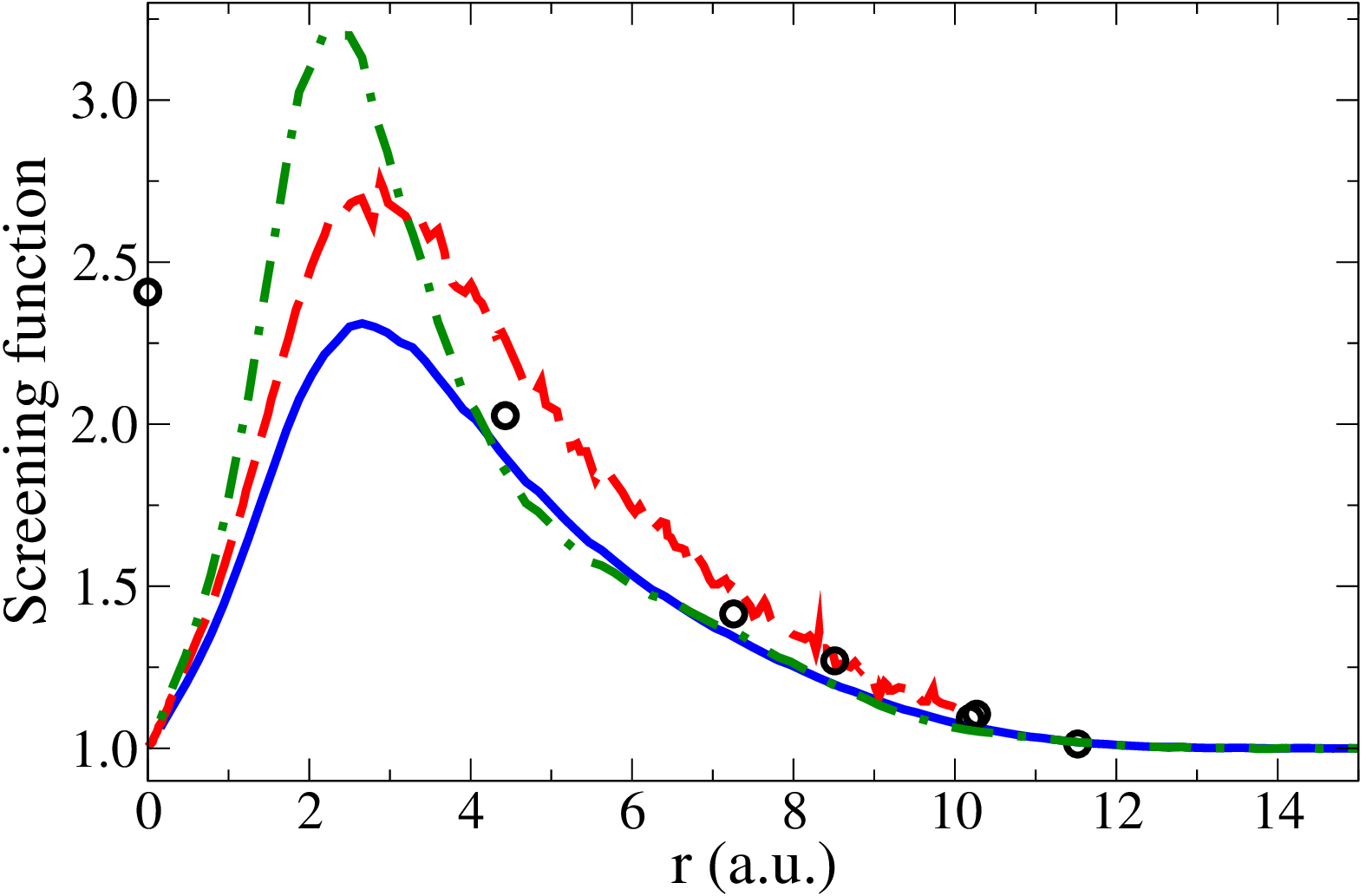}
\caption{\label{fig:fig3}(Color online) Spherical average of the screening dielectric function. Blue solid line, present calculation using an on-site parameter for Si U$_{at}$ = 10.6 eV; green dot-dashed line, present calculation using for Si U$_{at}$ = 8.8 eV; red dashed line, \textit{ab initio} calculation of Ref. \onlinecite{ogut03}; black circles, self-consistent tight-binding calculation of Ref. \onlinecite{delerue03}. }
\end{figure}

The localized orbital formalism described so far is particularly suitable for the study of point charge screening. The large interest in this problem is clearly related to the study of the excitonic interaction energy, in which a calculation of the full dielectric function is required.\cite{ogut03} Such a calculation is very demanding from a computational point of view,\cite{sottile05} and both models and semiempirical tools are often used in order to simplify the problem.\cite{ninno06}
From the definition of dielectric function, the screened electron-electron potential energy is given by 
\begin{equation}
 w \left(  \mathbf{r},\mathbf{r'},\omega \right) = \int \mathbf{dr_1} \epsilon ^{-1} \left( \mathbf{r},\mathbf{r_1}, \omega\right) v_b \left(  \mathbf{r_1},\mathbf{r'}\right).
\end{equation}
where the bare Coulomb interaction is $v_b (\mathbf{r},\mathbf{r'}) = e^2 / |\mathbf{r'}-\mathbf{r}|$.
According to Ref. \onlinecite{ogut03}, a common way to reduce the computational cost of the excitonic energy is by introducing a screening function, defined as
\begin{equation}
 \tilde{\epsilon}^{-1} \left(  \mathbf{r},\mathbf{r'},\omega \right) = \frac{ w \left(  \mathbf{r},\mathbf{r'},\omega \right) }{v_b \left(  \mathbf{r},\mathbf{r'}\right) }.
\end{equation}
Within the present scheme, using the equations previously introduced for the dielectric matrix, the following expansion is obtained:
\begin{equation}
 \tilde{\epsilon}^{-1} \left(  \mathbf{r},\mathbf{r'},\omega \right) = 1 + \frac{1}{e^2}\left|  \mathbf{r'}-\mathbf{r} \right| \sum _{\lambda,\mu} J _\lambda \left(\mathbf{r}\right) S _{\lambda,\mu} \left(\omega\right)  J _\mu\left(\mathbf{r'}\right).
\end{equation}
The screening function represents the dielectric response to a point charge placed at position $\mathbf{r}$. A common approximation consists of using a spherical average of the screening function calculated at $\mathbf{r}=0$ even for the response to an off-center point charge. The error made in the calculation of self-energies and excitonic energies can be often neglected.\cite{allan95,ogut03} So an effective screening function can be defined as $\bar{\epsilon} \left(  \mathbf{r},\omega \right) = 1/ \tilde{\epsilon}^{-1} (  \mathbf{0},\mathbf{r},\omega )$, whose spherical average is
\begin{equation}
 \bar{\epsilon} \left(r,\omega \right) = \frac{1}{1 + \frac{1}{e^2}r \sum _{\lambda,\mu} J _\lambda \left(\mathbf{0}\right) S _{\lambda,\mu} \left(\omega\right)  J _\mu\left(r\right)}.
\end{equation}
By using the approximations described in Sec. \ref{subsec:linear}, the Coulomb integral is reduced to the spherical average of the function $J_\lambda(\mathbf{r}) =J(\mathbf{r-R})$. We calculate $J$ using the Herman Skillman numerical tables for the atomic wave functions.\cite{herman63} With respect to previous tight-binding calculations, the present approach allows for a description of the screening in the whole real space.
The effective screening function for Si$_{35}$H$_{36}$ is shown in Fig. \ref{fig:fig3}. We report the results obtained  using different values of the on-site Coulomb interaction parameter $U_{at}$ for Si atoms. In one case, we used the value $U_{at}=8.8$ eV (green dot-dashed line), this value being calculated according to the prescription given above, using Eq. (\ref{eq:Aaverage}). In the second case, the results are obtained with $U_{at}=10.6$ eV (blue solid line), a value calculated as average between all the Coulomb energies between pairs of atomic orbitals.\cite{allan95,delerue97,delerue03} The present results are in reasonable agreement with both \textit{ab initio} calculations\cite{ogut03} (red dashed line) and self-consistent empirical tight-binding method\cite{delerue03} (symbols). It is important to remark that the peak position calculated here has the same value as in the \textit{ab initio} calculations. However, the present approach slightly underestimates the screening function in the space region close to the surface, and this is likely to be due to the approximations made and discussed above. It is worth to point out that the behavior of the screening function near the surface is independent of $U_{at}$, since it only depends on the dielectric constant of the material.

\begin{figure}
\includegraphics[width=0.45\textwidth]{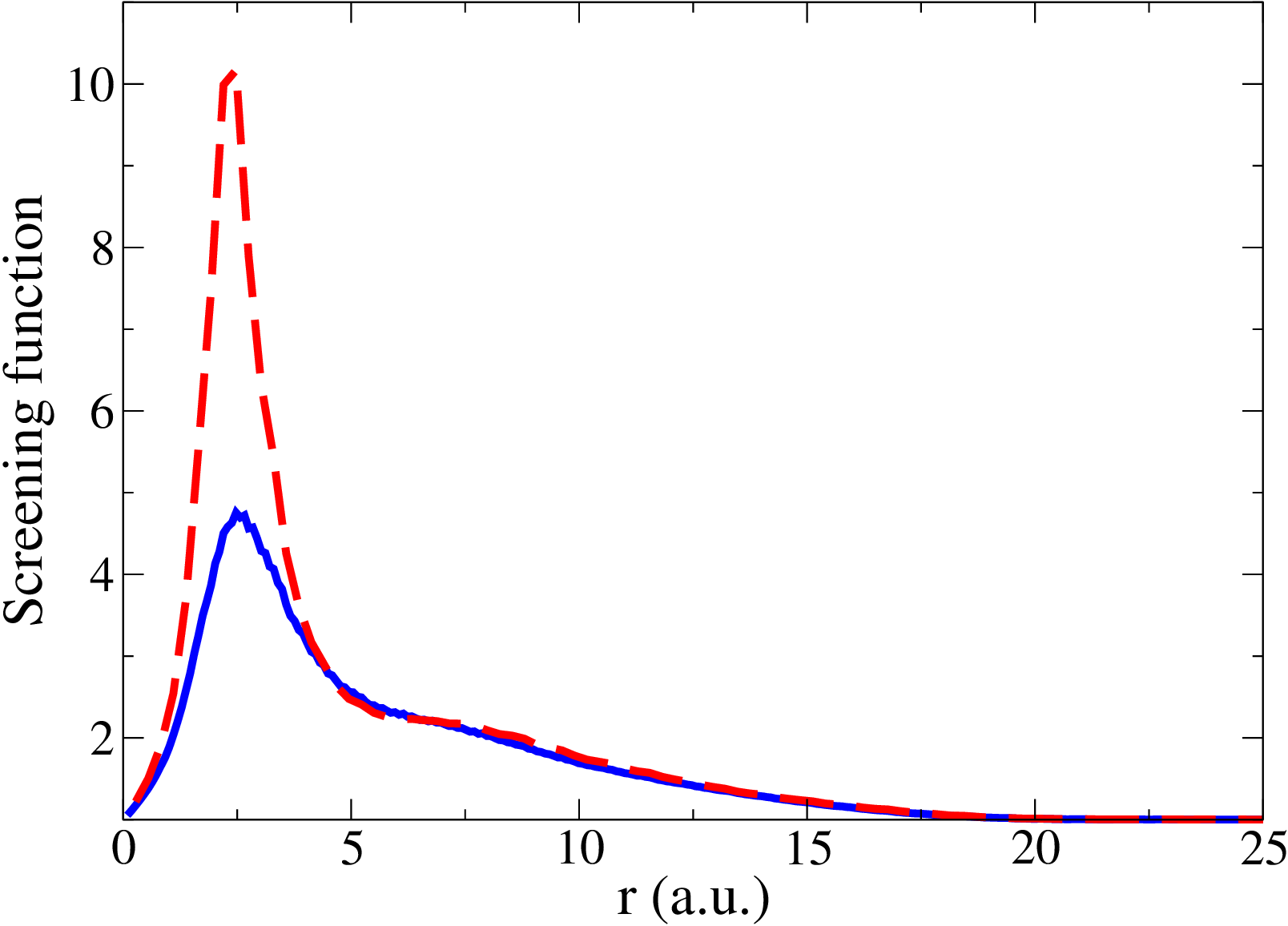}
\caption{\label{fig:fig4}(Color online) Spherical average of the effective screening function of Si$_{191}$H$_{148}$. Blue solid line, static screening ($\omega=0$); red dashed line, optical screening (calculated at the gap energy).}
\end{figure}

In Fig. \ref{fig:fig4}, the effective screening function for Si$_{191}$H$_{148}$ is reported. With respect to the smaller Si$_{35}$H$_{36}$, there is an increase of the screening, that is higher on increasing the size, in agreement with recent \textit{ab initio} calculations.\cite{franceschetti05} The static screening (blue solid line) is here compared with the optical screening, calculated at a frequency corresponding to the gap energy, that for Si$_{191}$H$_{148}$ is $E_0=2.53$ eV (red dashed line). The difference with respect to the static case is in the increase of the maximum, although the peak position remains unchanged. This is consistent with the fact that the real part of the dielectric constant at the gap energy has a larger value than in the static limit.

\begin{figure}
 \includegraphics[width=0.45\textwidth]{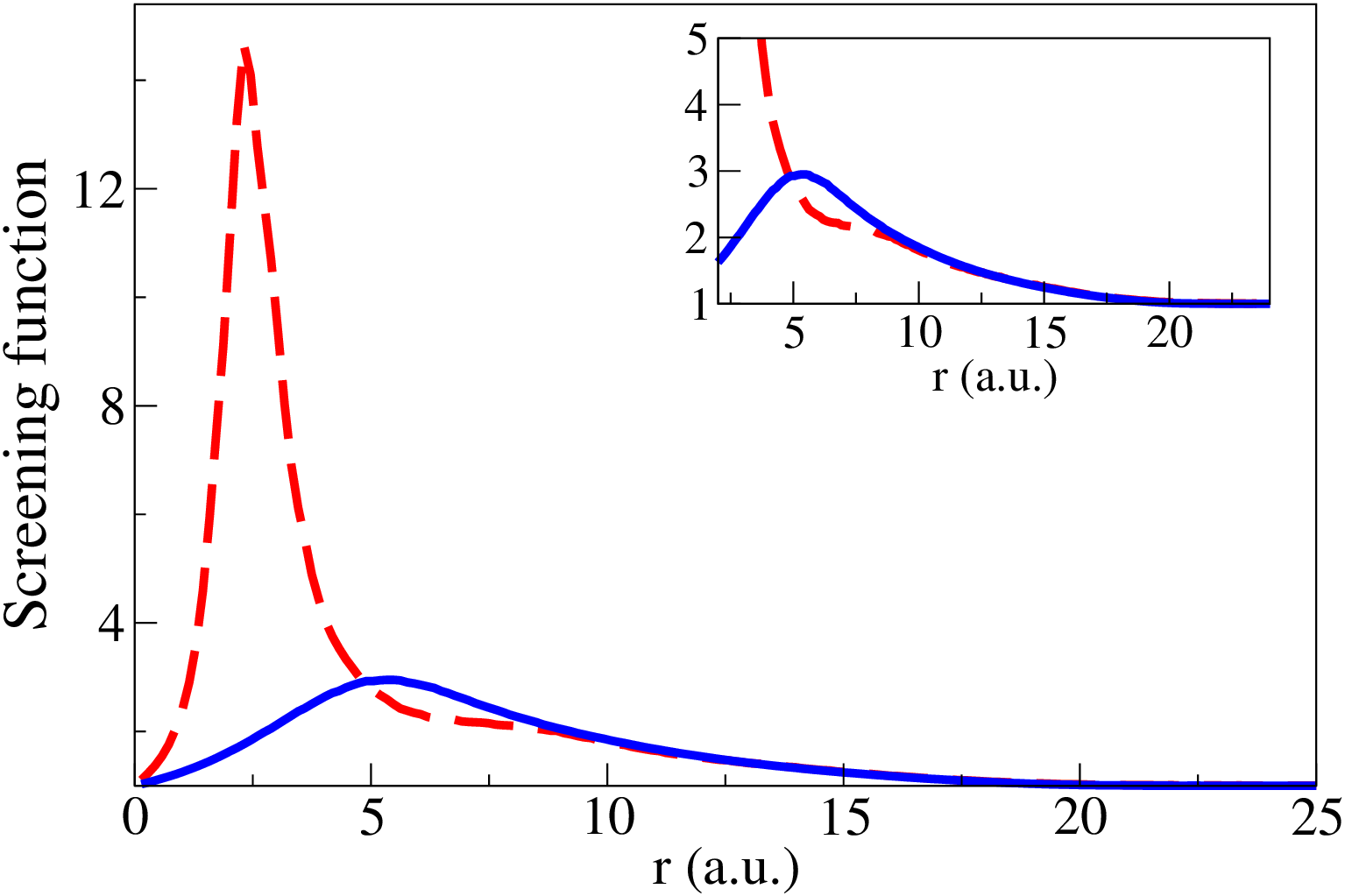}
\caption{\label{fig:fig5}(Color online) Screening function of a 2.0 nm diameter InAs nanocrystal. We studied the configuration with an As (In) atom placed in the nanocrystal center, indicated with the red dashed (blue solid) line. A zoom of the graph is shown in the inset.}
\end{figure}

In Fig. \ref{fig:fig5}, the effective screening function for a 2.0 nm size InAs nanocrystal is shown. As said above, this is only the electronic contribution to the screening, since we consider the atomic nuclei frozen at their bulk positions. The on-site Coulomb interaction parameters calculated according to the prescriptions given in Sec. \ref{subsec:linear} are U$_{at}$ = 9.68 and 6.25 eV for As and In atoms, respectively. At variance with Si, in the case of InAs nanocrystals there are two possible configurations for each size, depending on whether an indium (blue solid line) or an arsenic atom (red dashed line) is located at the center of the nanocrystal. As shown in Fig. \ref{fig:fig5}, both the intensity and the peak position assume different values for the two configurations. However, far from the impurity the screening function is exactly the same. This apparently strange result can easily be explained using a Thomas-Fermi model.\cite{ninno06,trani06} According to it, the point charge screening is composed of two contributions. A classical contribution, close to the surface, is due to the surface polarization charge induced by an external test charge located at the nanocrystal center. It is well described by classical electrostatics and only depends on the nanocrystal dielectric constant and radius. Indeed, it is significant that both the configurations show the same behavior close to the surface, due to the fact that they have the same macroscopic dielectric constant.
Approaching to the impurity site, the deviation from the classical model becomes important because of short range interactions. At variance with the classical contribution, the behavior in the neighborhood of the impurity (in particular, the peak position and intensity) strongly depends on the local environment around the impurity site and this is the cause of the difference between the two curves in Fig. \ref{fig:fig5}.
Therefore, InAs nanocrystals show important bulklike local field effects due to the difference of the atomic cores.
An interesting point confirming this view is that the surface polarization contribution does not depend on the on-site Coulomb term $U_{at}$. This parameter influences the amplitude of the screening function around its maximum, but it does not change the classical part close to the surface. In the inset the detail of the merging of the two curves is shown. The point at which this occurs can be indicatively defined as the screening radius of the nanocrystal.\cite{ninno06} 

\section{Conclusions}
In this paper, a tight-binding framework for the study of the dielectric response in confined systems has been described. The interest of the method consists in the easy inclusion of local field effects into the theory. We have shown that since local fields contribute significantly to the optical and screening properties of semiconductor nanocrystals, the use of an atomistic quantum mechanical framework has become absolutely necessary for the study of nanosized structures. The present framework is flexible, allowing the study of both optical and screening properties. It is computationally light and it gives results comparable to more sophisticated \textit{ab initio} methods with good agreement with experimental data. The method has been applied to the study of the optical and screening properties of both indirect gap (silicon) and direct gap (InAs) nanocrystals, with results that can be useful for deeper analysis. Moreover, by using the present framework, we demonstrate that the continuum dielectric model, almost always used to take into account local fields, is not adequate to study the electronic properties of very small nanocrystals.

\section*{ACKNOWLEDGMENTS}
This research is supported by the Italian government through the COFIN-PRIN 2005 project and the S.Co.P.E. project. All the calculations have been performed at CINECA-``Progetti Supercalcolo 2006'' and ``Campus Computational Grid''-Universit\`a di Napoli ``Federico II'' advanced computing facilities.


\begin{thebibliography}{56}
\bibitem[{not()}]{note}
\bibinfo{note}{In Ref. \onlinecite{trani07prb}, we named the CM with either the
  bulk Si or the RPA dielectric constant as classical and semiclassical model,
  respectively.}
\end{thebibliography}

\end{document}